\PassOptionsToPackage{table}{xcolor}
\documentclass[sigconf, nonacm]{acmart}
\AtBeginDocument{%
  }

\copyrightyear{2026}
\acmYear{2026}
\setcopyright{cc}
\setcctype{by}
\acmConference[CAIN '26]{2026 IEEE/ACM 5th International Conference on AI Engineering - Software Engineering for AI}{April 12--13, 2026}{Rio de Janeiro, Brazil}
\acmBooktitle{2026 IEEE/ACM 5th International Conference on AI Engineering - Software Engineering for AI (CAIN '26), April 12--13, 2026, Rio de Janeiro, Brazil}
\acmPrice{}
\acmDOI{10.1145/3793653.3793770}
\acmISBN{979-8-4007-2475-6/2026/04}

\usepackage{array}

\begin{document}

\title{Statistical Confidence in Functional Correctness: An Approach for AI Product Functional Correctness Evaluation}








\author{Wallace Albertini}
\email{wboquimpani@inf.puc-rio.br}
\orcid{0009-0002-0175-953X}
\affiliation{%
  \institution{Pontifical Catholic University of Rio de Janeiro}
  \city{Rio de Janeiro}
  \state{Rio de Janeiro}
  \country{Brazil}}

\author{Marina Condé Araújo}
\email{maraujo@inf.puc-rio.br}
\orcid{0009-0003-3374-898X}
\affiliation{%
  \institution{Pontifical Catholic University of Rio de Janeiro}
  \city{Rio de Janeiro}
  \state{Rio de Janeiro}
  \country{Brazil}}

\author{Júlia Condé Araújo}
\email{jcaraujo@inf.puc-rio.br}
\orcid{0009-0009-8787-8062}
\affiliation{%
  \institution{Pontifical Catholic University of Rio de Janeiro}
  \city{Rio de Janeiro}
  \state{Rio de Janeiro}
  \country{Brazil}}

\author{Antonio Pedro Santos Alves}
\email{apsalves@inf.puc-rio.br}
\orcid{0000-0002-3415-6201}
\affiliation{%
  \institution{Pontifical Catholic University of Rio de Janeiro}
  \city{Rio de Janeiro}
  \state{Rio de Janeiro}
  \country{Brazil}}

\author{Marcos Kalinowski}
\email{kalinowski@inf.puc-rio.br}
\orcid{0000-0003-1445-3425}
\affiliation{%
  \institution{Pontifical Catholic University of Rio de Janeiro}
  \city{Rio de Janeiro}
  \state{Rio de Janeiro}
  \country{Brazil}}

\renewcommand{\shortauthors}{Albertini et al.}

\begin{abstract}
The quality assessment of Artificial Intelligence (AI) systems is a fundamental challenge due to their inherently probabilistic nature. Standards such as ISO/IEC 25059 provide a quality model, but they lack practical and statistically robust methods for assessing functional correctness. This paper proposes and evaluates the Statistical Confidence in Functional Correctness (SCFC) approach, which seeks to fill this gap by connecting business requirements to a measure of statistical confidence that considers both the model's average performance and its variability. The approach consists of four steps: defining quantitative specification limits, performing stratified and probabilistic sampling, applying bootstrapping to estimate a confidence interval for the performance metric, and calculating a capability index as a final indicator. The approach was evaluated through a case study on two real-world AI systems in industry involving interviews with AI experts. Valuable insights were collected from the experts regarding the utility, ease of use, and intention to adopt the methodology in practical scenarios. We conclude that the proposed approach is a feasible and valuable way to operationalize the assessment of functional correctness, moving the evaluation from a point estimate to a statement of statistical confidence.
\end{abstract}




\maketitle

\section{Introduction}

Artificial Intelligence (AI) systems are rapidly evolving and are responsible for decisions that directly affect human lives, ranging from medical diagnoses and autonomous driving to financial forecasting and judicial support systems. However, this pervasive integration shed light on the importance of safety, trustworthiness, accountability, and reliability when developing such systems \cite{Nguyen2024, simonetta2024iso}. A recent review of AI systems revealed that about a quarter of companies reported facing failure rates of up to 50\% in AI projects \cite{oveisi2024review}. There is an urgent demand for robust regulatory frameworks and evaluation criteria, especially given the high social-economic costs associated with such failures \cite{moor2012turing}. 

A core challenge in evaluating such systems lies in a paradigmatic incompatibility, as the application of deterministic testing methodologies to systems that are, by nature, probabilistic and adaptive. AI models do not conform to the predictable input–output mappings typical of traditional software systems, rendering conventional testing approaches largely inadequate. This inadequacy is compounded by the well-known “test oracle problem” \cite{barr2014oracle}, which in the context of AI models acquires unique complexity, since the “correct” or expected outputs are often undefined, uncertain, or only approximable within a probabilistic range \cite{aleti2023software}. Moreover, the black-box nature of many algorithms, coupled with the immaturity of current quality assurance processes and tools for AI, further exacerbates the difficulty of assessing correctness, robustness, and bias.

In response to these emerging issues, the international standardization community has sought to adapt existing quality models and create new ones capable of addressing the specific challenges posed by AI systems. Among these initiatives, the ISO/IEC 25059 standard stands out as a significant development \cite{ISO25059}, representing an adaptation of the ISO/IEC 25010 quality model to the context of AI~\cite{ISO25010, OVIEDO2024StandardIA}.

Although some methodological and technological frameworks have been proposed for measuring AI system quality and performance \cite{oviedo2024environment}, the literature still lacks empirical and industrial case studies that explore the practical difficulties of conducting such evaluations. In practice, product evaluation challenges are still latent, particularly regarding the definition of specification limits and the selection of representative and meaningful test cases~\cite{verdugo2024connecting}.

To address these gaps, this study proposes the \textit{Statistical Confidence in Functional Correctness} (SCFC) approach, which integrates statistical sampling and inference principles with capability analysis for the assessment of functional correctness in AI systems. The approach consists of four steps: defining quantitative specification limits with the stakeholders, collecting a representative data sample via stratified sampling with the support of domain experts, applying bootstrapping to estimate a confidence interval for the performance metric, and calculating a capability index as
a final indicator to analyze if the AI product delivers acceptable functional correctness.

To evaluate the approach, we conducted a case study involving two AI products of different companies. We applied the SCFC approach to evaluate the functional correctness of both products. Thereafter, we presented the results of these cases to four experienced AI practitioners (one from each company and two from completely independent contexts) and conducted semi-structured interviews to gather feedback on the suitability of the approach and its acceptance. We found that the experts considered the proposed approach suitable and robust, and showed overall acceptance. Furthermore, they provided useful insights into the applicability and prerequisites of each individual step for some specific AI contexts. These insights provide practical recommendations, and we detail them as part of our results.

Based on our evaluations, we conclude that the SCFC approach and the recommendation described in this paper help to provide practical means for the systematic evaluation of functional correctness of AI systems, offering a suitable way for confronting stakeholder specification limits with the non-deterministic nature of intelligent systems.

The remainder of this paper is organized as follows. Section \ref{sec:background} presents the background and related work. Section \ref{sec:approach} describes the proposed sampling-based approach. Sections \ref{sec:case_study_design} and \ref{sec:case_study_results} present the case study design and results. Sections \ref{sec:discussion} and \ref{sec:threats} bring the discussion and threats to the validity of our study. Finally, Section \ref{sec:conclusion} contains the concluding remarks.

\section{Background and Related Work} \label{sec:background}

The application of AI systems in production environments still crawls towards maturity. Less than 50\% of AI projects reach production \cite{zimelewicz2024}. There are various challenges regarding transitioning traditional software engineering to AI-based systems \cite{martinez2022, nahar2023meta, KALINOWSKI2025107866}.

One of the foundational pillars for software quality evaluation is the ISO/IEC 25000 family of standards (SQuaRE) \cite{ISO25000}. Recently, ISO/IEC 25059 \cite{ISO25059} was released in order to adapt the product quality model defined in ISO/IEC 25010 and focus on the specific characteristics of AI systems.
Although ISO/IEC 25059 represents an important starting point, the standard remains high-level. It still requires defining concrete metrics and thresholds to be effectively applied in industrial contexts. 

\citet{oviedo2024environment} addressed this gap by proposing a methodological and technological environment for evaluating Functional Suitability in AI systems, aligned with ISO/IEC 25059. To assess the sub-characteristic of \textit{Functional Correctness}, the authors define the property Functional Accuracy, which verifies whether the success rate of a requirement meets or exceeds a pre-established threshold. To determine the number of test executions required for this measurement, they employ a statistical formula to calculate the sample size, considering a given acceptable error level.

Although the approach proposed by \citet{oviedo2024environment} represents a significant advancement in the operationalization of the standard, it focuses on classifying a requirement as “accurate” or “inaccurate” based on a fixed threshold. While useful, this assessment does not quantify model performance variability nor provide a measure of statistical confidence in the observed result. There remains a clear need to adapt existing metrics and develop new methods and tools to measure AI-specific sub-characteristics.




\section{The Statistical Confidence in Functional Correctness Approach}\label{sec:approach}

\begin{figure}[h]
  \centering
  \includegraphics[width=\linewidth]{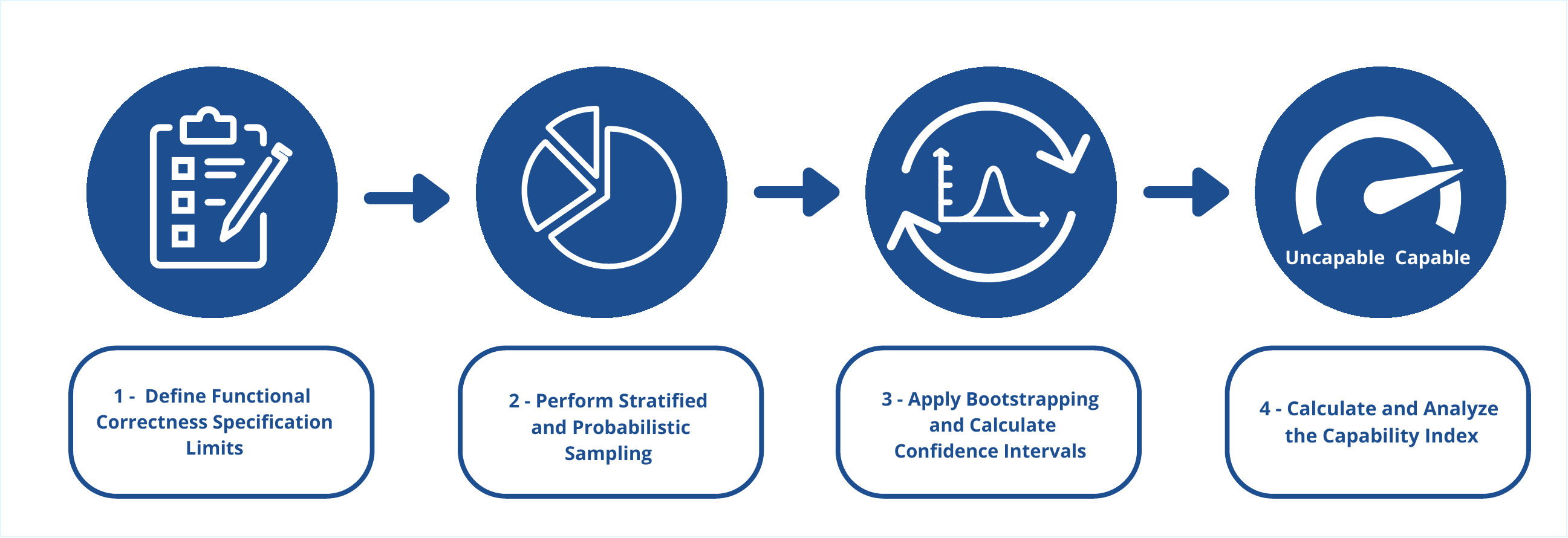}
  \caption{Steps of the SCFC approach.}
  \label{fig:flowchart}
\end{figure}

To address the need for a more robust evaluation of functional correctness in AI systems, this paper proposes a structured methodological approach. The objective is to overcome the limitations of evaluations based solely on point estimates (such as a single accuracy average), which may mask performance variability and create a false sense of confidence. The proposed methodology incorporates statistical uncertainty into the evaluation process to generate a capability index that quantifies the real confidence in meeting business requirements. The approach unfolds in four sequential steps, as presented in Figure ~\ref{fig:flowchart}.

\subsection{Define Functional Correctness Specification Limits}

The foundation of any quality assessment relies in the clarity of what is considered “acceptable”. Therefore, the first step of our approach consists of translating a business need into an acceptance criterion that is objective, unambiguous, and measurable. This stage is crucial to connect technical evaluation with business value and to avoid ambiguous judgments based on vague or implicit requirements.

The main action here is the formal establishment of specification limits for an acceptable functional correctness. For a given performance metric (such as recall, precision, F1-score, or a custom business metric), a Lower Specification Limit (LSL) is defined, representing the minimum acceptable performance. For example, the business need “the fraud detection system must not miss many fraudulent cases” can be translated into the technical requirement: “The system must achieve a fraud recall of at least 95\% to be deployed in production.” This formalizes the requirement, establishing LSL = 0.95 and a clear target for evaluation.

On the other hand, an Upper Specification Limit (USL) is defined by metrics where lower values are preferable, such as the false positive rate. In the same hypothetical example of a fraud detection system, we could have the technical requirement: "The system must have a false positive rate below 10\%." This formalizes the USL = 0.1.

\subsection{Perform Stratified and Probabilistic Sampling}

The validity of any software test, particularly for AI systems, depends intrinsically on the quality and representativeness of the data used \cite{clemmensen2022data}. Using a test sample that does not reflect the real-world data distribution encountered in production can lead to misleading conclusions. To mitigate this risk, this stage focuses on collecting a statistically robust test sample.

The approach employs stratified sampling, a technique superior to simple random sampling in scenarios with heterogeneous subpopulations. The process begins with the identification, in collaboration with a domain expert, of stratification variables, those data characteristics that may influence model behavior (e.g., transaction type, geographic region, user profile). Next, based on historical data or business knowledge, the real proportion of each stratum in the population is determined. Finally, to avoid selection bias, elements within each stratum are probabilistically selected. This method ensures that even minority but potentially critical subgroups are adequately represented in the test sample, resulting in a more reliable and generalizable evaluation.

\subsection{Apply Bootstrapping and Calculate Confidence Intervals}

The probabilistic nature of AI systems means that their performance on a specific test sample is merely an estimate of their true performance. Repeating the test with a different sample would likely yield a slightly different result. To capture this variability, the third stage of our approach moves beyond point estimation and focuses on estimating the performance distribution.

\begin{figure}[h]
  \centering
  \includegraphics[width=\linewidth]{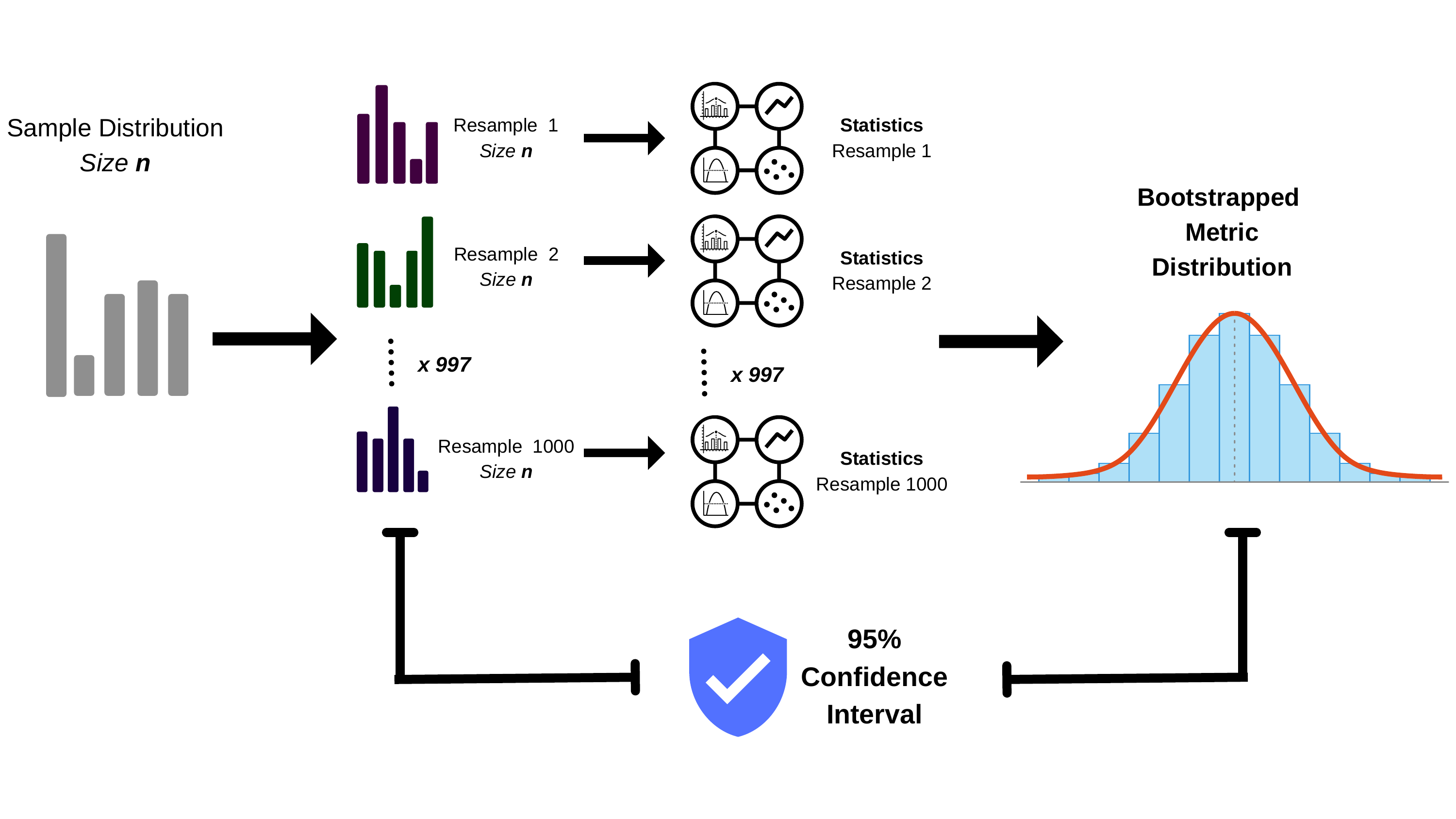}
  \caption{Bootstrapping procedure.}
  \label{fig:bootstrapping}
\end{figure}

To this end, we apply the non-parametric resampling technique known as Bootstrapping to the collected test sample. This technique has been reported to be more reliable than simple inference statistics from one sample\cite{bootstrapping}. This method simulates the collection of multiple new samples (with replacement) of size \(n\) from the original test set, bootstrapping the metrics of interest for each of the \(S\) resamples. We assume \(n\) as our original collected test sample \cite{bootstrappingPopulationSizeParam}, and we set 1000 for \(S\), which is a value that is reported to allow meaningful statistics \cite{bootstrappingReplacementParam}. The result is an empirical distribution of the metrics of interest, which allows us to compute a Confidence Interval (CI), typically at the 95\% level. Figure \ref{fig:bootstrapping} summarizes this bootstrapping procedure.

The CI provides a range of plausible values for the system’s true performance. For example, a mean recall of 89\% is a limited piece of information. However, a 95\% CI between [87\% - 91\%] informs us, with high statistical confidence, that the true system performance lies within that range. This information is more valuable, as it quantifies uncertainty and supports a more informed decision considering the required specification limit defined in Step 1.

\subsection{Calculate and Analyze the Capability Index}

The final stage of the approach synthesizes the previous information into a single indicator to measure how capable the system is of consistently meeting the functional correctness specification limit. To this end, we propose a capability index that adapts the standard Six Sigma $C_{pk}$ formulation\cite{yang2003design}. The underlying principle remains to assess the relationship between the average performance and its variability relative to the specification limit.

The key innovation of our approach lies in how variability is measured. While traditional $C_{pk}$ uses the standard deviation, often assuming normality of the data, our methodology employs a capability measure derived directly from the CI obtained via bootstrapping. This measure, corresponding to the distance between the mean and the relevant CI bound, is non-parametric and therefore more robust for performance metric distributions of AI systems, which are not necessarily normal.

\textbf{Capability Index for the Lower Limit ($C_{pl}$):}

Measures the system’s capability relative to the LSL:

\[
C_{pl} = \frac{\text{Mean} - \text{LSL}}{\text{Mean} - \text{Lower bound of CI}}
\]

\textbf{Capability Index for the Upper Limit ($C_{pu}$):}

Measures the system’s capability relative to the USL:

\[
C_{pu} = \frac{\text{USL} - \text{Mean}}{\text{Upper bound of CI} - \text{Mean}}
\]

The final system capability index, $C_{pk}$, is then defined as the minimum of the two values. This ensures that the assessment always reflects the worst-case performance scenario, that is, the system’s capability is dictated by the specification limit most at risk of being violated.

\[
C_{pk} = \min(C_{pu}, C_{pl})
\]

The application of this formula is straightforward:
\begin{itemize}
    \item If a requirement has only a LSL (e.g., recall $>$ 95\%), only $C_{pl}$ is calculated and $C_{pk} = C_{pl}$ (the $C_{pu}$ is considered infinite).
    \item If a requirement has only an USL (e.g., error rate $<$ 2\%), only $C_{pu}$ is calculated and $C_{pk} = C_{pu}$ (the $C_{pl}$ is considered infinite).
    \item If a requirement has both limits (e.g., latency between 100ms and 200ms), both indices are computed, and $C_{pk}$ is the smaller of the two.
\end{itemize}

Regardless of the scenario, the interpretation of the final $C_{pk}$ value remains consistent and decision-oriented:

\begin{itemize}
    \item $C_{pk} < 1.0$ (Unacceptable): The capability is lower than the safety margin. The system does not consistently meet the specification.
    \item $C_{pk} = 1.0$ (Minimum Capability): The capability is approximately equal to the safety margin. The system is at the limit of its capability and may be considered for deployment in non-critical cases, but requires careful monitoring.
    \item $C_{pk} > 2.0$ (Excellent): The capability is at least twice the distance from the safety margin, indicating very high confidence that the system will consistently meet the specification.
\end{itemize}

\section{Case Study Design}\label{sec:case_study_design}

To evaluate the applicability and perceived value of the proposed approach in a practical context, we designed an exploratory case study. The study aims to evaluate the SCFC approach through its application in real-world AI systems and to gather qualitative perceptions from industry experts. Following the guidelines for case study research by \citet{runeson2012case}, this section details the goal, research questions, case selection, and data collection and analysis procedures.

\subsection{Goal and Research Questions}

The goal of this study follows the GQM (Goal-Question-Metric) goal definition template, and is defined as follows \cite{van2002goal}: 
\textbf{Analyze} the SCFC approach \textbf{for the purpose of} characterization \textbf{with respect to} perceived suitability for assessing functional correctness, perceived utility, ease of use, and intention to adopt, \textbf{from the point of view of} experts involved in engineering AI systems and assessing software product quality, \textbf{in the context of} two real-world AI systems.

From this goal, we derived the following Research Questions (RQ) to guide the study:

\textbf{RQ1: How suitable is the proposed approach, and each of its steps, for assessing the functional correctness of AI systems?} To address this research question, we created subquestions for each step.

\begin{itemize}
    \item \textbf{RQ1.1:} Is the definition of functional correctness specification limits perceived as suitable to establish acceptance criteria?
    \item \textbf{RQ1.2:} Is stratified and probabilistic sampling perceived as suitable to ensure the representativeness of data?
    \item \textbf{RQ1.3:} Is the use of bootstrapping and confidence intervals perceived as suitable to understand system performance variability?
    \item \textbf{RQ1.4:} Is the capability index perceived as suitable to support deployment decisions?
\end{itemize}

\textbf{RQ2: What is the level of acceptance of the proposed approach among experts?} To address this research question, we created subquestions for each construct of the technology acceptance (TAM) model~\cite{davis1989perceived}.

\begin{itemize}
    \item \textbf{RQ2.1:} What is the perceived usefulness of the approach in improving confidence in AI quality assessments?
    \item \textbf{RQ2.2:} What is the perceived ease of use when applying the approach within their current evaluation processes?
    \item \textbf{RQ2.3:} What is the experts’ intention to adopt or recommend this approach in future projects?
\end{itemize}

\subsection{Case and Context Selection}

Two real-world AI systems from distinct domains were selected to enhance the generalizability of the results. The selection targeted cases representative of common challenges in AI functional correctness evaluation.

 \subsubsection{Case 1: Cargo Deck Space Estimation AI System}

\begin{figure}[h]
  \centering
  \includegraphics[width=\linewidth]{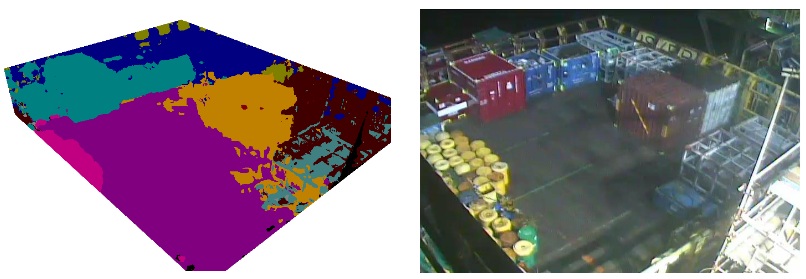}
  \caption{STEGO image segmentation of a real oil platform deck, with purple pixels representing the deck floor.}
  \label{fig:modeck}
\end{figure}

The first selected case study is a cargo deck space estimation AI system, Figure ~\ref{fig:modeck}, whose purpose is to analyze images of the cargo deck on offshore oil platforms to estimate the available free space. The optimization of space allocation on offshore platforms represents a logistical challenge in which inefficient resource management can lead to operational delays, high costs, and safety risks. The system was developed to address this problem, representing a typical AI system used in real business scenarios, where success directly impacts logistical efficiency.

From a technical perspective, this system combines two machine learning models: STEGO (Self-supervised Transformer with Energy-based Graph Optimization), responsible for performing semantic segmentation, classifying each pixel in the image as belonging to one of two categories: “deck floor” or “object”, and KNN (k-Nearest Neighbors), which converts the count of pixels classified as “deck floor” into a real-world area estimate, expressed in square meters (m²). This model was trained to account for camera perspective distortion, where pixels farther from the field of view correspond to larger real-world areas.

For evaluation purposes, the functional correctness criterion was defined as follows: for each image, an acceptance range (composed of a lower and upper limit) for the free-space estimate is manually calculated by a domain expert. The business requirement, which translates into our LSL, was defined as follows: the system would be considered functionally correct if its predictions fell within the acceptance range for at least 70\% of the test images. This quantifiable requirement makes this system a suitable case study for our approach, enabling the direct application of the capability index calculation to evaluate whether the system meets this criterion with statistical confidence.

 \subsubsection{Case 2: Credit Card Fraud Detection System}

The second selected case study is a machine learning model for detecting fraud in credit card transactions. The purpose of this system is to analyze transactions in real-time to classify each one as "legitimate" or "fraudulent". Payment fraud is a critical problem for the financial industry, resulting in significant financial losses for consumers and institutions, as well as undermining trust in the digital payment ecosystem. This model represents a crucial, automated line of defense, where high performance and reliability are essential for risk mitigation and customer protection.

From a technical perspective, the system is a binary classification model. It was trained on a vast dataset of historical transactions, using a variety of features such as transaction amount, geographical location, purchase frequency, and the cardholder's history to learn the patterns that distinguish fraudulent activities from genuine transactions. A central challenge in developing this model is the severe class imbalance, where fraudulent transactions are extremely rare events compared to legitimate ones, requiring specific modeling and evaluation techniques.


For evaluation purposes, we considered what was truly fraud and what the model inferred as fraud, and generated a confusion matrix, which is described in Table \ref{tab:confusionmatrix}. The correctness criterion focuses on the recall metric, which measures the proportion of actual frauds that the model correctly identifies. Given the high cost of a False Negative (a missed fraud), the business requirement, which translates into our LSL, was defined as follows: the system would be considered functionally adequate if it correctly identified at least 98\% of the actual fraudulent transactions in the test set. This rigorous and quantifiable requirement, centered on a critical business metric, makes this model an ideal case study for our approach, enabling the direct application of the capability index calculation to evaluate whether the system meets this criterion with statistical confidence.

\begin{table}[h]
\centering
\caption{Confusion Matrix}
\label{tab:confusionmatrix}
\begin{tabular}{|c|c|c|}
\hline
 & \multicolumn{2}{c|}{\textbf{Model Inferred Fraud}} \\ \hline
\textbf{Fraud} & No & Yes \\ \hline

No  & 109,269,288 (True Negative) & 10,365 (False Positive) \\ \hline
Yes & 82 (False Negative)         & 8,951 (True Positive) \\ \hline

\end{tabular}
\end{table}

\subsection{Data Collection Procedures}

\subsubsection{Calculation of metrics}

The proposed approach integrates the concept of bootstrapped confidence intervals and process capability indices to quantify both performance and uncertainty. To calculate the metrics for each selected case, initially, a specification limit was defined with the company stakeholders to represent the minimum acceptable performance threshold aligned with business or operational requirements. 

After the specification limit was established, we identified a representative sample for each selected case with the support of domain experts from the company. Thereafter, we applied bootstrapping with resampling to generate 1000 simulated samples, allowing the estimation of the sampling distribution of the performance metric.

From this distribution, a 95\% confidence interval was derived, establishing the range within which the true performance of the system is expected to lie with high statistical confidence. This measure allowed to assess not only whether each system met its target performance but also how stable and reproducible that performance was across different data samples.

Finally, the capability index (Cpk) was calculated to quantify each system’s ability to consistently satisfy the specified limits. This index combines both the average performance and the dispersion of the results, providing an interpretable indicator of robustness. Values of Cpk greater than 1.0 generally indicate that the process variability remains within acceptable margins, supporting informed decisions regarding deployment readiness.

\subsubsection{Interviews}

To evaluate the proposed approach from the perspective of industry professionals, after applying it to both selected AI projects, we conducted a series of semi-structured interviews, which are appropriate for investigating how individuals experience a phenomenon~\cite{robson2024real}. The objective was to foster an in-depth discussion about the perceived benefits, limitations, and applicability of the method for assessing functional correctness in AI systems.

A semi-structured interview is a qualitative research method that uses a script of open-ended questions as a guide while offering the researcher the flexibility to explore interesting responses and delve deeper into topics that emerge during the conversation~\cite{robson2024real}. Unlike rigid questionnaires, this approach allows for capturing the complexity and rationale behind participants' perceptions, generating rich and detailed data based on their professional experiences~\cite{given2008sage}. In software engineering research, interviews are widely used to gather feedback from experts on innovative practices, tools, and methods, as well as to understand the challenges of the industrial context~\cite{runeson2012case}.

\begin{figure}[h]
  \centering
  \includegraphics[width=\linewidth]{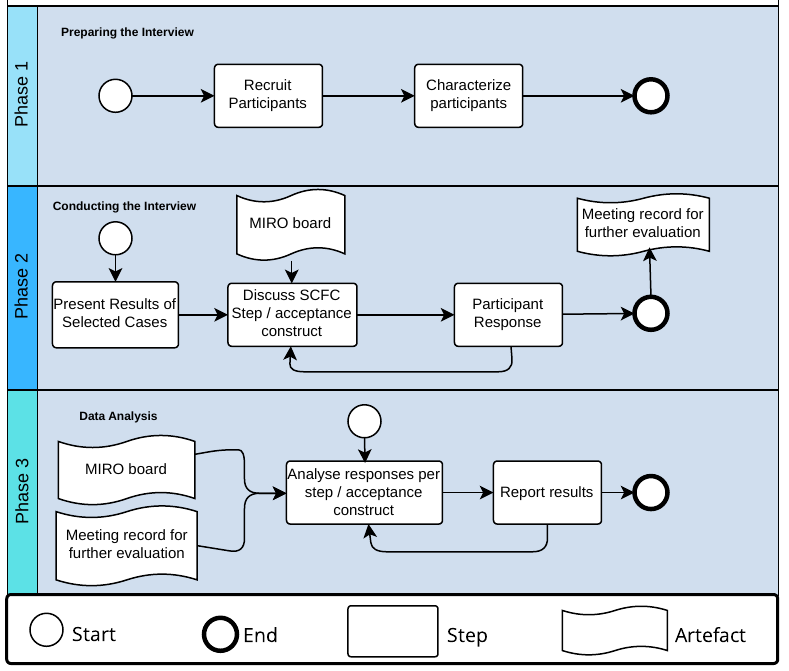}
  \caption{Interview process overview.}
  \label{fig:Interview}
\end{figure}

\begin{figure}[h]
  \centering
  \includegraphics[width=\linewidth]{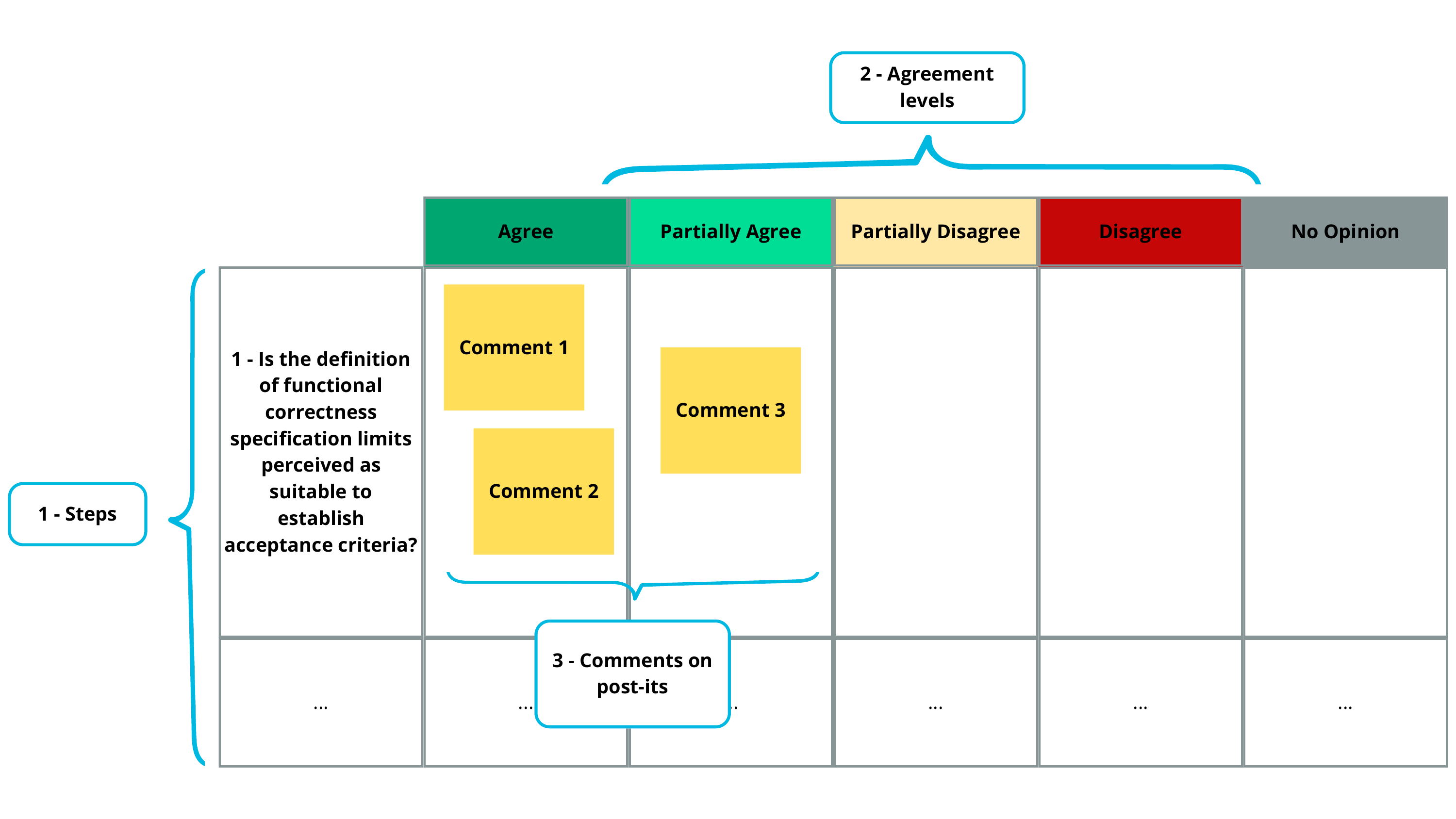}
  \caption{Miro board semi-structured interview template.}
  \label{fig:miro}
\end{figure}

The interview process was structured into three main phases to ensure systematic data collection and analysis, as illustrated in Figure ~\ref{fig:Interview}. These phases are described hereafter: 

\begin{itemize}

    \item \textbf{Prepare the semi-structured interviews:} This phase involved the recruitment and characterization of participants using purposive sampling. We selected four AI specialists, all with extensive experience in both industrial and academic AI projects, as detailed in the Table \ref{tab:exemplo}. 
    
    To ensure a contextually grounded analysis, we recruited the lead professionals responsible for each selected case study (P2 and P4). We clarify that these participants were the same experts who provided the datasets and project definitions. While this ensures deep domain knowledge, we acknowledge the potential for positive bias. To mitigate this and assess generalizability, we recruited two additional independent professionals (P1 and P3) with no prior link to the projects or the data. These external participants were selected based on their solid expertise—possessing 8 and 6 years of experience, respectively—to judge the methodology's utility across different domains. They were presented with the results of the case studies to evaluate the proposed approach.
    
    \item \textbf{Conducting the semi-structured interviews}: This phase consisted of conducting the interviews. The researcher presented the results of applying SCFC to both selected cases (15 minutes) and then asked the participant questions about the suitability of each SCFC step and about the TAM acceptance constructs (perceived usefulness, perceived ease of use, and intention to adopt), using a Miro board for them to take post-it notes on the responses (\textit{cf} Figure \ref{fig:miro}) using a Likert scale \cite{joshi2015likert}. The entire interview was also recorded for further analysis.
   
    \item \textbf{Analyze data and results}: The final phase involved the joint interpretation of results from the Likert scale assessments and the qualitative analysis. This integration enabled the identification of acceptance patterns and insights regarding the perceived applicability and usefulness of the proposed approach for evaluating functional correctness in AI systems.

\end{itemize}

\begin{table*}[h]
\centering
\caption{Characterization of the Participants}
\label{tab:exemplo}
\begin{tabular}{|p{1cm}|p{2.5cm}|p{2.8cm}|p{2.5cm}|p{2.5cm}|p{2.5cm}|}
\hline
\textbf{ID} & \textbf{Education level} & \textbf{Years of experience} & \textbf{Role} & \textbf{Industry} & \textbf{Company size} \\ \hline
P1& Phd. & 8 years & Research engineer & Software R\&D & ~200 \\ \hline
P2& Masters Degree & 2 years & ML engineer & Oil and gas & ~50  \\ \hline
P3& Masters Degree & 6 years & AI engineer & Finance & ~5000 \\ \hline
P4& Phd. & 13 years & Staff Data Science & Payments & ~10000 \\ \hline
\end{tabular}
\end{table*}

\subsection{Analysis Procedures}

To answer the research questions, we combined the quantitative data analysis of applying our approach to both selected cases with the qualitative analysis of the experts' perceptions.

\textbf{Quantitative Data Analysis:} This step focused on the data generated by applying the SCFC approach to the two selected cases. For each case, the Cpk was calculated as described in Section 3. The analysis consisted of interpreting the Cpk value against the predefined thresholds (<1.0 Unacceptable, =1.0 Minimum Capability, >2.0 Excellent). The outcome of this quantitative analysis served as the primary objective evidence of each system's capability and as a starting point for the discussion with the experts.


\textbf{Qualitative Data Analysis:} The qualitative data, collected from the semi-structured interviews, were analyzed to answer RQs 1 and 2. The interview recordings were transcribed and subjected to open coding, where relevant concepts and opinions were identified. Regarding the analysis procedure and quality control, the initial coding was performed by one researcher. To mitigate individual bias and improve the reliability of the findings, a second senior researcher reviewed the extracted codes. Instead of calculating statistical agreement rates, disagreements were resolved through consensus meetings until full agreement was reached on the interpretation of the findings. Additionally, Likert scale responses were aggregated to quantify the perceived suitability of each step of the approach (RQ1). The TAM questionnaire responses were analyzed to measure perceived usefulness, perceived ease of use, and intention to adopt (RQ2).

\section{Case Study Results}\label{sec:case_study_results}

This section presents the main findings from the case study conducted to evaluate the applicability and perceived value of the proposed approach for measuring functional correctness in AI systems. The analysis provides perceptions of how the methodology performs in real-world contexts and how it is interpreted by experienced practitioners.

\subsection{Applying the Approach}

In this section, we present the results of applying the SCFC approach to both selected cases. The objective is to demonstrate the step-by-step methodology and how its results support more informed decision-making about functional correctness and deployment.

\subsubsection {Evaluation of the Cargo Deck Space Estimation AI System}

The evaluation followed the four steps of the SCFC approach:

\begin{enumerate}
    \item \textbf{Define Functional Correctness Specification Limits:} The starting point was the LSL established by the stakeholders: the system should correctly classify (\textit{i.e.}, with the prediction falling within the manual acceptance range) at least 70\% of the test images (LSL = 0.70).

    \item \textbf{Perform Stratified and Probabilistic Sampling:} The system was run on a test set of 42 real-world images of cargo decks that, according to domain experts, were representative of typical operational conditions. The observed performance was 35 correct predictions, resulting in a prediction acceptance rate of approximately 83\% (35/42). An analysis of the incorrect predictions revealed that the failures occurred in images that presented visual challenges, such as shadow, fog on the camera, and/or containers with deck floor color, which hindered correct segmentation by the model. Although this point value is above the requirement, it does not offer insight into the variability or statistical confidence of the result.

    \item \textbf{Apply Bootstrapping and Calculate Confidence Intervals:} To quantify uncertainty, we applied bootstrapping analysis to the sample results. The mean of the resulting bootstrap distribution was 0.8340. More importantly, the analysis generated a 95\% CI for the prediction acceptance rate, which was [0.7143, 0.9286]. This means that with 95\% confidence, the true performance of the cargo deck space estimation system on the data population lies within this range.

    \item \textbf{Calculate and Analyze the Capability Index:} With the LSL, the mean, and the lower bound of the CI, we calculated the Capability Index ($C_{pl}$), which in this case corresponds to the final $C_{pk}$, as there is no upper specification limit.
    
    \[
    C_{pk} = \frac{(\text{Mean} - \text{Lower Bound of CI})}{(\text{Mean} - \text{LSL})} \approx 1.12
    \]
    
    The result of the Capability Index ($C_{pk}$) for the cargo deck space estimation system was 1.12.
\end{enumerate}

A $C_{pk}$ of 1.12 places the system above the required minimum capability. This classification reveals an important nuance that the 83\% average alone cannot express: while the average performance (83.4\%) is comfortably above the 70\% requirement, the system's variability, as reflected in the confidence interval, shows that the lower bound of its performance (71.43\%) is very close to the specification limit (70\%). This indicates that the system's safety margin is not wide.

Based on this analysis, the decision was that the system \textbf{"can be considered for deployment."} This recommendation, however, would be accompanied by the recommendation that continuous monitoring in production is advisable to ensure its performance does not degrade and violate the minimum requirement. The approach, therefore, not only validated the system but also qualified the risk associated with its deployment.

\subsubsection {Evaluation on the Fraud Detection Model}

Following the same structure, the approach was also applied to the second selected case, the machine learning model for detecting credit card transaction fraud. The objective was to quantitatively assess whether the model's performance meets the rigorous quality criteria required for its deployment. The evaluation followed the same four steps of the SCFC approach:

\begin{enumerate}
    \item \textbf{Define Functional Correctness Specification Limits:} The starting point was the LSL defined by stakeholders. Due to the high criticality of failing to detect a fraud (a False Negative), the requirement was established based on the recall metric: the model should correctly identify at least 98\% of fraudulent transactions (LSL = 0.98).

    \item \textbf{Perform Stratified and Probabilistic Sampling:} The model was run on a test dataset, and its results were consolidated in the previously presented confusion matrix \ref{tab:confusionmatrix}. The observed performance was 8,951 True Positives and 82 False Negatives, resulting in a recall of approximately 99.1\% ($8951 / (8951 + 82)$). Although this value is above the requirement, it offers no insight into the variability or statistical confidence of the result.

    \item \textbf{Apply Bootstrapping and Calculate Confidence Intervals:} To quantify uncertainty, we applied bootstrapping analysis to the sample results. The mean of the resulting bootstrap distribution was 0.9911, confirming the initial estimate. More importantly, the analysis generated a 95\% CI for the recall, which was [0.9855, 0.9967]. This means that, with 95\% confidence, the true performance of the model on the data population lies within this range.

    \item \textbf{Calculate and Analyze the Capability Index:} With the LSL, the mean, and the lower bound of the CI, we calculated the Capability Index ($C_{pk}$), using the same approach as in the previous case.
\end{enumerate}

\[
C_{pk} = \frac{\text{Mean} - \text{LSL}}{\text{Mean} - \text{Lower CI Bound}} \approx 1.98
\]




The result of the Capability Index ($C_{pk}$) for the fraud detection model was 1.98. This places the model near the excellent category. Regarding the ground truth, this model was considered suitable for deployment and is currently running in a production environment, as is the system from the first case study.

We highlight that a direct reliability comparison between the two systems is not intended, as they operate in different contexts and evaluate different metrics with distinct specification limits. Instead, the approach ensures that each system is reliable within its own scope. In this specific case, the $C_{pk}$ of 1.98 reveals a robust situation: the lower bound of performance (98.55\%) is comfortably above the specification limit (98\%), quantifying the high confidence in the model’s adherence to its specific requirements.

\subsection{Interview discussions}

The semi-structured interviews were conducted with four industry experts experienced in AI engineering, one working on each of the selected cases (P2 on the cargo deck space estimation system and P4 on the credit card fraud detection system), and two from independent contexts (P1 and P3). The average duration of the interviews was 41 minutes. The objective was to analyze qualitative feedback to answer our two main research questions (RQ1 and RQ2), focusing on the perceived suitability of the approach and its level of acceptance. The following sections analyze the experts' responses to each research question in detail, synthesizing the points of consensus and the main recommendations identified.

\subsection{RQ1. How suitable is the proposed approach, and each of its steps, for assessing the functional correctness of AI systems?}

Overall, the experts considered the SCFC approach and its four steps suitable for evaluating functional correctness in AI product evaluations. The agreement levels on the suitability of each step are shown Table~\ref{tab:interview_steps_all}. However, the experts provided nuances that provide additional details on the applicability and prerequisites of each individual step. Hereafter, we elaborate on the feedback received for each SCFC step.

\begin{table}[htbp]
\centering
\caption{Agreement on the suitability of SCFC's steps}
\label{tab:interview_steps_all}
\setlength{\tabcolsep}{3pt}
\renewcommand{\arraystretch}{1.00}
\scriptsize
\begin{tabular}{|>{\centering\arraybackslash}m{0.9cm}|c|c|c|c|c|c|c|c|c|c|c|c|c|c|c|c|}
\hline
\cellcolor{gray!40} & \multicolumn{4}{c|}{\textbf{Step 1}} & \multicolumn{4}{c|}{\textbf{Step 2}} & \multicolumn{4}{c|}{\textbf{Step 3}} & \multicolumn{4}{c|}{\textbf{Step 4}} \\ \hline
\cellcolor{gray!40} & \textbf{P1} & \textbf{P2} & \textbf{P3} & \textbf{P4}
& \textbf{P1} & \textbf{P2} & \textbf{P3} & \textbf{P4}
& \textbf{P1} & \textbf{P2} & \textbf{P3} & \textbf{P4}
& \textbf{P1} & \textbf{P2} & \textbf{P3} & \textbf{P4} \\ \hline
\cellcolor{green!40}\textbf{Strongly Agree}&  & X &  & X  & X  &   & X & X & X & X  & X & X  &   & X  &  X &  X \\ \hline
\cellcolor{green!15}\textbf{Partially Agree}& X &   & X &  &  & X &  &   &  &  &   &   & X  &   &   &   \\ \hline
\cellcolor{orange!15}\textbf{Partially Disagree}& &   &   &   &   &   &   &   &   &   &   &   &   &   &   &   \\ \hline
\cellcolor{red!40}\textbf{Strongly Disagree}& &   &   &   &   &   &   &   &   &   &   &   &   &   &   &   \\ \hline
\cellcolor{gray!70}\textbf{No Opinion}& &   &   &   &   &   &   &   &   &   &   &   &   &   &   &   \\ \hline
\end{tabular}
\end{table}

\subsubsection{RQ1.1: Is the definition of functional correctness specification limits perceived as suitable to establish acceptance criteria?}

There was a consensus among the experts that establishing quantitative specification limits is an essential practice, aligned with business and statistical needs to define a clear baseline for the evaluation. It is noteworthy that the participants involved in the two selected cases (P2 and P4) were particularly convinced by the practical utility of defining specification limits in real-world scenarios, potentially due to having seen the practical utility in projects in which they were directly involved. Three important attention points were raised during the interviews:

\begin{itemize}
    \item \textbf{Stakeholder Dependency:} The effectiveness of this step directly depends on the knowledge of domain experts or Product Owners (POs). P4 emphasized the need for "input from business stakeholders." P1 expanded on this limitation, questioning the step's feasibility in "problems that are too abstract," where not even stakeholders might know or be able to define a clear threshold.
    \item \textbf{Scope of the Approach:} P3 (who partially agreed) noted that the selected cases focused on classification problems, which limited his visibility on the approach's application in more complex contexts, such as multimodal models or systems with multiple AI components.
    \item \textbf{Single vs. Multiple Metrics:} P3 and P4 suggested that the approach, as presented, focuses on a single-dimension metric. P4 suggested that if there are multiple metrics of interest, they should be mapped to a single 1-dimensional metric that contains all the business meaning. However, the approach is not intended to be limited to a single metric, as several different metrics (and inferences on their variations) can be calculated based on bootstrapping.
\end{itemize}

\subsubsection{RQ1.2: Is stratified and probabilistic sampling perceived as suitable to ensure the representativeness of data?}

The experts agreed (P1, P3, and P4) that stratified sampling is a feasible and valuable technique, as it "simulates the production environment" (P3) and it is necessary to "ensure the correct representation when 100\% sampling is impossible" (P4). 

However, two conditions for its proper application were highlighted:

\begin{itemize}
    \item \textbf{Quality of the Original Dataset:} P1 pointed out that the technique assumes the original dataset is already representative of reality. Applying stratified sampling to a dataset with "low representativeness" of the real world will not correct the fundamental bias.
    \item \textbf{Representativeness vs. Case Need:} P2 (who partially agreed) brought a nuance: proportional representativeness is not always desirable. In cases like fraud detection, it might be interesting to over-represent (oversample) the minority class (fraud) due to its "dynamic and changing nature." Also, it was mentioned governance and fairness principles, where it is fundamental to avoid the under-representation of ethnic minorities so as not to perpetuate biases.
\end{itemize}

\subsubsection{RQ1.3: Is the use of bootstrapping and confidence intervals perceived as suitable to understand system performance variability?}

This step had the strongest and most unanimous agreement among the experts. All participants agreed on the suitability of using bootstrapping and CIs. P4 highlighted that bootstrapping is the "standard methodology to empirically determine the distribution of an estimator" when the real distribution is unknown. P2 complemented this, stating that the technique allows "the measurement of data variability" and "the generalization of statistical results". P1 also explicitly stated his agreement, "I completely agree", and P3 classified the technique as "robust for what it proposes".

\subsubsection{RQ1.4: Is the capability index perceived as suitable to support deployment decisions?}

The Capability Index was generally well-received but generated more debate about its added value. P2 considered it a robust indicator, a "summary metric that directly communicates system performance", highlighting its usefulness for "comparison between models" in an MLOps scenario. The main considerations were:

\begin{itemize}
    \item \textbf{Value in Non-Marginal Cases:} P1 questioned how much the index adds in extreme cases. According to him, if the difference between the mean and the specification limit is very large (positive or negative), the decision to deploy or not would already be "obvious" without the index, leading him to ask: "to what extent is it worth introducing another metric?". However, we highlight that the Capability Index is also worthy for cases where model performance variation can be extremely high and that risks might still exist even if the mean is far from the specification limit.
    \item \textbf{Dependency on the Metric:} P4 reinforced that the index's validity depends on it being applied to a "single metric that has to be optimized" and that captures the full business meaning. However, in some cases, different metrics might complement each other.
\end{itemize}

\subsection{RQ2. What is the level of acceptance of the proposed approach among experts?}

The general acceptance level of the approach was high, as shown in Table~\ref{tab:interview_tam_all}. The experts perceived SCFC as useful (PU), easy to use (PEU), and showed a behavioral intention (BI) to adopt it, motivated mainly by filling a gap in current QA processes. Hereafter, we further elaborate on each TAM construct.

\begin{table}[h!]
\centering
\caption{TAM responses in the interviews}
\label{tab:interview_tam_all}
\setlength{\tabcolsep}{3pt}
\renewcommand{\arraystretch}{1.00}
\scriptsize
\begin{tabular}{|>{\centering\arraybackslash}m{1.5cm}|c|c|c|c|c|c|c|c|c|c|c|c|}
\hline
\cellcolor{gray!40} & \multicolumn{4}{c|}{\textbf{PU}} & \multicolumn{4}{c|}{\textbf{PEU}} & \multicolumn{4}{c|}{\textbf{BI}} \\ \hline
\cellcolor{gray!40} & \textbf{P1} & \textbf{P2} & \textbf{P3} & \textbf{P4}
& \textbf{P1} & \textbf{P2} & \textbf{P3} & \textbf{P4}
& \textbf{P1} & \textbf{P2} & \textbf{P3} & \textbf{P4} \\ \hline
\cellcolor{green!40}\textbf{Strongly Agree}& X  & X &  X & X  & X  & X  &   &  X & X & X  & X &  X \\ \hline
\cellcolor{green!15}\textbf{Partially Agree}&  &   &  &   &  &  & X &   &  &  &   &   \\ \hline
\cellcolor{orange!15}\textbf{Partially Disagree}& &   &   &   &   &   &   &   &   &   &   &   \\ \hline
\cellcolor{red!40}\textbf{Strongly Disagree}& &   &   &   &   &   &   &   &   &   &   &   \\ \hline
\cellcolor{gray!70}\textbf{No Opinion}& &   &   &   &   &   &   &   &   &   &   &   \\ \hline
\end{tabular}
\end{table}

\subsubsection{RQ2.1: What is the perceived usefulness of the approach in improving confidence in AI quality assessments?}

The usefulness of the approach was strongly recognized. P3 stated that it proposes "another way to ensure that an AI product can be successful in production, using more robust statistical metrics". P2 and P4 highlighted its practical utility in MLOps, as it "helps create continuous monitoring" (P4) and "can be used in different contexts" (P2). P1 added that having confidence intervals "helps define stakeholder expectations".

\subsubsection{RQ2.2: What is the perceived ease of use when applying
the approach within their current evaluation processes?}

Most experts considered the approach simple to apply. P1 stated, "I don't see any complexity," and P2 highlighted that the "calculations are simple" and have "negligible computational costs." The only barrier to use identified was the bootstrapping step. P3 (who partially agreed) mentioned that practitioners without prior contact with the technique might face a "learning curve." P4, although considering it simple, noted that bootstrapping "might require a separate process to be implemented, as it involves more computation", in contrast to the other steps. However, it is noteworthy that bootstrapping does not require re-executing the model inferences, but only resampling examples and their already processed inference results.

\subsubsection{RQ2.3: What is the experts’ intention to adopt or recommend this approach in future projects?}

There was a positive consensus regarding the intention to adopt. P3 was positive on the adoption, stating that the approach "shows great promise in helping practitioners" fill the gap we see today in evaluation techniques, confirming his intention to adopt. P1 also stated: "Yes, I would like to use it." P2 saw the approach as an excellent "guideline for building a framework for continuous analysis." Finally, P4 supported its adoption to "bring the discussion of quality thresholds into the process" of development and to ensure good methodologies.

\section{Discussion}\label{sec:discussion}

The central objective of this work was to propose and evaluate a practical and statistically robust approach for assessing the functional correctness of AI systems, filling a gap in the operationalization of standards such as ISO/IEC 25059~\cite{ISO25059}. The validation of the SCFC approach, through its application to real-world AI systems and the collection of expert perceptions, suggests that it is both feasible to apply and valuable in practice.

The main result of our approach is the paradigm shift: from evaluating a point estimate metric (like an 83\% accuracy) to a Capability Index (Cpk = 1.12) that incorporates variability. The cargo deck space estimation case study perfectly exemplifies this transition: an 83\% result seems, at first glance, much higher than the 70\% requirement. However, the Cpk of 1.12 classifies barely above the minimum required capability, revealing that the safety margin is small in relation to the model's instability. This single number communicated a layer of risk that the mean alone concealed. The experts in our interviews validated this usefulness, highlighting that the use of confidence intervals helps to "define stakeholder expectations" (P1) and that the Cpk is a "summary measure that directly communicates performance" (P2).

The interviews were also crucial for identifying the prerequisites and boundary conditions of the approach. The first step, defining Lower and Upper Specification Limits (LSL/USL), was universally accepted as essential, but its effectiveness is entirely dependent on process maturity and stakeholder involvement. As pointed out by P1, the approach may face difficulties in "problems that are too abstract," where the limits are not well-known. Likewise, the stratified sampling step was considered robust, but its validity, as P2 recalled, can be situational; in domains such as fraud or those requiring fairness, proportional sampling may not be ideal, suggesting that the sampling technique must be adapted to the problem's context.

Regarding the capability index, a pertinent criticism was raised by P1 about its obviousness in cases where performance is clearly very good or very bad. While this observation is valid, we believe the value of Cpk lies not only in marginal cases but in its ability to standardize decision-making. In an MLOps and continuous monitoring environment, as mentioned by P2 and P4, having a standardized indicator that removes the subjectivity of an engineer's "feeling" about a metric is a significant process benefit.

A perceived limitation was the applicability to a single-dimension metric, as suggested by P3 and P4. In fact, many real-world AI systems require a balance (or trade-off) between multiple metrics (\textit{e.g.}, precision vs. recall, or performance vs. latency). However, we believe that this limitation concerns a misunderstanding, as there is no such limitation in the approach, and multiple metrics can be considered without problems. Of course, the interpretation of criticality could have to consider the metric with the lowest ($C_{pk}$).

Finally, the experts confirmed a high intention to adopt, viewing the approach as potential "guidance for building a continuous analysis framework" (P2) that fills a gap and "shows great promise in helping practitioners" (P3). Furthermore, the approach was considered easy to apply in practice, with the main learning curve being the application of bootstrapping (P3). These findings provide initial indications that position SCFC as a potentially useful and easily applicable approach for AI engineering teams to assess functional correctness.

\section{Threats to Validity}\label{sec:threats}

When conducting a qualitative and exploratory case study like the one presented in this paper, it is essential to acknowledge the potential threats to the validity of the results. We discuss the limitations of our study based on the four categories for threats to validity described by \citet{wohlin2024experimentation}.

\paragraph{Conclusion Validity} This validity refers to the ability to draw correct conclusions from the data. The main threat here is the limited number of participants. Our qualitative analysis is based on interviews with four experts. Although all are experienced and from different industrial contexts, their opinions may not represent the entire AI engineering community. Considering this, we understand the need for additional empirical evaluations and avoid overclaiming the validity of our results. Nevertheless, the strong agreement of the participants and the qualitative feedback allowed us to understand at least some situations in which the approach would be useful in practice. Hence, we believe that it is valuable to share the approach with the community.

\paragraph{Internal Validity} This validity concerns factors that may have influenced the interview results. There is a risk of researcher bias, as the authors themselves presented the approach to the experts before interviewing them. This may have introduced an acquiescence bias, where participants might tend to agree with the approach. We mitigated this by ensuring anonymity in responses and using open-ended and neutral questions, actively encouraging critical feedback (which was also received, as seen in the discussions about sampling and Cpk).

\paragraph{Construct Validity} This validity questions whether we are measuring what we intend to measure. Our research questions (RQ1 and RQ2) were designed to measure the constructs of "suitability," and "acceptance." The threat is that the interview questions may not have properly captured these concepts. To mitigate this, the questions about acceptance (RQ2) were inspired by the TAM technology acceptance model.
    
\paragraph{External Validity} This is, perhaps, the most significant threat and refers to the generalizability of the results. Our study is based on two selected cases and the perceptions of a small group of four experts from different industrial contexts. We cannot claim that our approach is universally applicable to all types of AI systems (such as multimodal models, as one expert pointed out) or that all practitioners will have the same perception. We mitigated this by selecting cases from distinct domains and experts with varied backgrounds. However, we acknowledge this limitation and position our results as exploratory preliminary evidence of the approach's feasibility and usefulness, which must be validated in future studies with more cases and contexts.

\section{Concluding Remarks}\label{sec:conclusion}

The evaluation of the functional correctness of AI systems, given their probabilistic nature, remains a significant challenge for software engineering. Standards like ISO/IEC 25059 point in the right direction, but the industry lacks practical methods for their implementation. This article sought to address this issue by proposing the SCFC approach, which connects business requirements for functional correctness to a statistical evaluation that quantifies risks of not achieving these requirements. 

Our main contribution is SCFC's four-step approach, involving defining specification limits, performing stratified and probabilistic sampling, applying bootstrapping and calculating confidence intervals, and calculating and analyzing a capability index. The innovative aspect of the approach concerns integrating inferential statistics into functional correctness software product quality evaluation. In doing so, it allows development teams to move the conversation from simple performance measurement to a more mature discussion about statistical confidence. 

Its application in selected industrial cases demonstrated its ability to generate a decision indicator that considers variability and quantifies deployment risk more effectively than a point-average metric. Furthermore, interviews with experts confirmed the perceived usefulness and low adoption barrier, and recognized the approach as filling a practical gap in AI product quality evaluation with respect to functional correctness. 

\section*{Artifacts Availability}
Artifacts related to applying the approach to the selected cases are private, given that they concern real projects. Thus, we provide a synthetic example of a classification problem that applies the approach and is runnable on Google Colab. Artifacts related to the interviews, including the Miro board with the post-its registered by the participants during the interviews and anonymized transcripts of the complete interviews, are available in our open science repository~\cite{zenodoRepo}.

\begin{acks}

We express our gratitude to the Brazilian Research Council - CNPq (Grant 312275/2023-4), Rio de Janeiro State's Research Agency - FAPERJ (Grant E-26/204.256/2024), the Coordination for the Improvement of Higher Education Personnel (CAPES), the  Kunumi Institute, and Stone Co. for their generous support.

\end{acks}


\bibliographystyle{ACM-Reference-Format}
\bibliography{sample-base}


\end{document}